\documentclass[]{aa}

\usepackage{graphicx}
\usepackage{txfonts}

\begin{document}

   \title{Identification of periodic density structures in Solar Orbiter data: Length scales and radial evolution}

  \author{ C. Katsavrias \inst{1,2}, S. Di Matteo \inst{1,3}, L. Kepko\inst{1}, N. M. Viall\inst{1}, \and  A. Walsh\inst{4} }

   \institute{NASA-Goddard Space Flight Center, Greenbelt, MD, USA
\\
              \email{ckatsavrias@gmail.com}
        \and
              Department of Physics, National and Kapodistrian University of Athens, Athens, Greece 
        \and
               Physics Department, The Catholic University of America, Washington, DC, USA
        \and
               ESA/ESAC, Madrid, Spain}

   \date{Received September 15, 1996; accepted March 16, 1997}

\titlerunning{PDS within SIR} 
\authorrunning{Katsavrias et al.}

  \abstract
   {The quasi-Periodic density structures (PDSs) are quasiperiodic variations of solar wind density ranging from a few minutes to a few hours. They are trains of advected density structures with radial length scales $L_R\approx$100–10,000 Mm, thus belonging to the class of solar wind “mesoscale structures”. Even though PDS at L1 have been extensively studied both through statistical and event analysis, their investigation at distances closer to the Sun is limited.  }
   {This study performs a statistical investigation of PDS at various distances from the Sun between 0.3 and 1 AU by exploiting Solar Orbiter data. }
   {We compiled and made publicly available an extensive list of PDSs following a well-established methodology that combines the Multitaper method as well as wavelet analysis to reveal the distribution of PDS radial length scales and how they vary with respect to the radial distance.}
   {Our results indicate that PDS advected with the ambient slow solar wind are expanded at a rate of approximately 10\%, while PDS detected during fast solar wind segments show compression at a similar rate.}
   {These are consistent with the scenario in which PDSs are formed at the Sun by processes involving magnetic reconnection and interchange reconnection in the solar corona. }


   \maketitle

\section{Introduction}

The quasi-periodic density structures (PDSs) are a subset of solar wind ``mesoscale structures'' \citep{Viall2021} and a particular and prevalent constituent of the ambient solar wind \citep{Kepko2020}. Their frequencies span the 0.1 to 5 mHz range, corresponding to advected periodic radial length scales of 100–10,000 Mm, and the train of PDSs can last several hours to over a day. While studied less, their azimuthal size scales have been estimated as $\approx$2000 Mm and $\approx$1000 Mm for PDSs with frequencies less and higher than 1 mHz, respectively \citep{SanchezDiaz2017,DiMatteo2024}. These structures can impact global magnetospheric \citep{Borovsky2006} and radiation belt dynamics because of the direct changes to the solar wind dynamic pressure, created by the changing proton number
density\citep{Viall2009a,DiMatteo2022a,Kurien2024}. Therefore, the investigation of their occurrence rate, generation mechanisms as well as their propagation is of great importance to understanding the daily driving of geospace and the wider space physics community.

The precise generation mechanisms of PDSs are still debated, but a consensus is that many are formed in the solar atmosphere, while others are newly created en route via processes like turbulence \citep{Verscharen2019,Viall2021}. For example, several remote-sensing observations of the Sun show direct evidence of PDSs of solar origin \citep{Viall2015,Ventura2023,Alzate2024}. These observations together with in situ observations of magnetic and elemental composition associated with PDSs, have provided evidence of solar wind released through magnetic reconnection of previously closed magnetic fields \citep[e. g.,][]{Antiochos2011,Reville2020,Kepko2024}. Further studies have also associated PDSs with small flux ropes \citep{Higginson2018,DiMatteo2019b,Lavraud2020,Lynch2023,Katsavrias2024}. Finally, PDS appear to travel with the solar wind flow \citep{Viall2015}. 

Even though PDS at L1 and in the near-Earth region are extensively studied both through statistical and event analysis, their investigation at distances closer to the Sun is limited. Analysis of in situ observations made by Helios at distances less than 0.5 AU have also identified PDSs \citep{DiMatteo2019b} and possible candidates in similar structures \citep{Stansby2018,Berriot2024,Wallace2025}. In this work, we perform--for the first time to our knowledge--an extensive statistical study to identify PDSs in Solar Orbiter data spanning the 0.3 to 1 AU radial distance from the Sun. To that end, we use a well-established methodology combining the Multitaper method and wavelet analysis to accurately detect PDSs both in the frequency and time domains. The manuscript is organized as follows: section 2 describes the data as well as the particular technique we used to identify PDSs. In section 3 we present the results of the PDS frequency and radial length scale distributions. Finally, we discuss our results and summarize our conclusions in section 4.

\section{Data and methods}

We used the total ion distributions as well as solar wind speed from the Proton and Alpha particle Sensor \citep[PAS;][]{Owen2020} on board the Solar Orbiter spacecraft, spanning the 2021--2024 time period. The resolution of the PAS data is $\approx$ 4 s and for this study we have reduced the cadence to 20 s. 

To identify PDS in the Solar Orbiter density time series, we followed a spectral analysis procedure based on the multitaper method \citep[MTM;][]{Thomson1982,DiMatteo2021}. First, we estimated the adaptive MTM power spectral density over running linearly de-trended six-, twelve- and eighteen-hour intervals with a 5 minutes sliding window. Then, we identified periodic variations in the time series at frequency values at which both the normalized spectrum and an additional statistical test for phase coherence (harmonic F-test) exceed the corresponding 90\% confidence threshold. The combination of the MTM and the harmonic F-test is a robust approach that has been extensively tested and employed to identify PDSs in in situ measurements \citep{Viall2008,Viall2009a,Viall2009b,DiMatteo2019a,Kepko2020,Kepko2024} and remote-sensing observations \citep{Viall2010,Viall2015}. Here, we implement the more recent version of this approach \citep[SPD\_MTM;][see also appendix \ref{MTMmethod}]{DiMatteo2020_software}, which was employed in earlier investigations of PDSs in solar wind plasma, composition, and remote sensing measurements \citep{DiMatteo2022a,Gershkovich2022,Gershkovich2023,Alzate2024}.

\begin{figure}
{\includegraphics[width=\hsize]{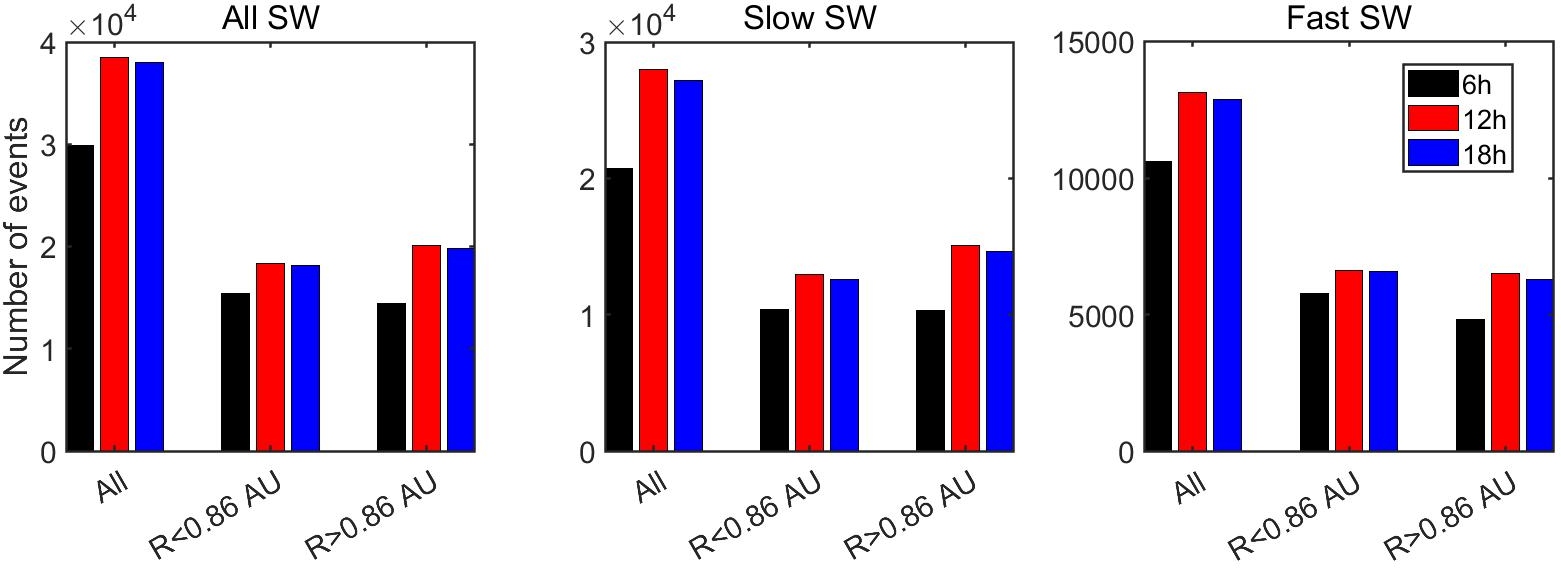}}     
\caption{Distribution of the number of PDS events. From left to right: distributions for the entire sample, for the slow and fast solar wind, respectively. The distributions of the six-, twelve- and eighteen-hour interval are coloured with black, red and blue, respectively.}
\label{Fig1}
\end{figure}

In order to determine the exact duration of each PDS event, we used the continuous wavelet transform \citep[CWT;][]{Torrence1998} and the Morlet wavelet \citep{Morlet1982} as mother wavelet, following \citet{Katsavrias2024,Katsavrias2025}. We calculated the wavelet power of the total ion density time series in each of the six-, twelve- and eighteen-hour intervals identified as containing a PDS via the MTM method. Then, we isolated the time periods in which the power at the corresponding PDS frequency and at each time step exceeded the 75th percentile of the power during the entire six-, twelve- and eighteen-hour interval, again at the same frequency identified with the MTM method. Note that we examined the average power in the $\pm0.15$ mHz range in order to account for the frequency uncertainty. Finally, in order to avoid any dubious detections, we required the duration of each PDS to be at least twice its period. An example of the aforementioned event detection procedure is shown in figure \ref{FigA1} in the appendix. This criterion produced a list of 29891, 38478 and 37974 PDS events (left panel in figure \ref{Fig1}) for the six-, twelve- and eighteen-hour interval, respectively, which are publicly available \citep{Katsavrias&DiMatteo2025}. These events were further divided into slow and fast solar wind (middle and right panels in figure \ref{Fig1}), corresponding to velocities below and above 450 km/s, respectively. Moreover, we divided the events with respect to the spacecraft radial distance (R). We note that due to the nature of the spacecraft orbit, Solar Orbiter spends much more time near-L1 than in distances closer to the Sun. Consequently, our PDS list contains fewer events as we move to lower radial distances (see also figure \ref{FigA2} in the appendix). Therefore, the R$=$0.86 AU limit was chosen in order to retain approximately the same number of events at near-L1 and inner distances. 

\section{Results}

\subsection{Distribution of PDS frequencies}

\begin{figure}
{\includegraphics[width=\hsize]{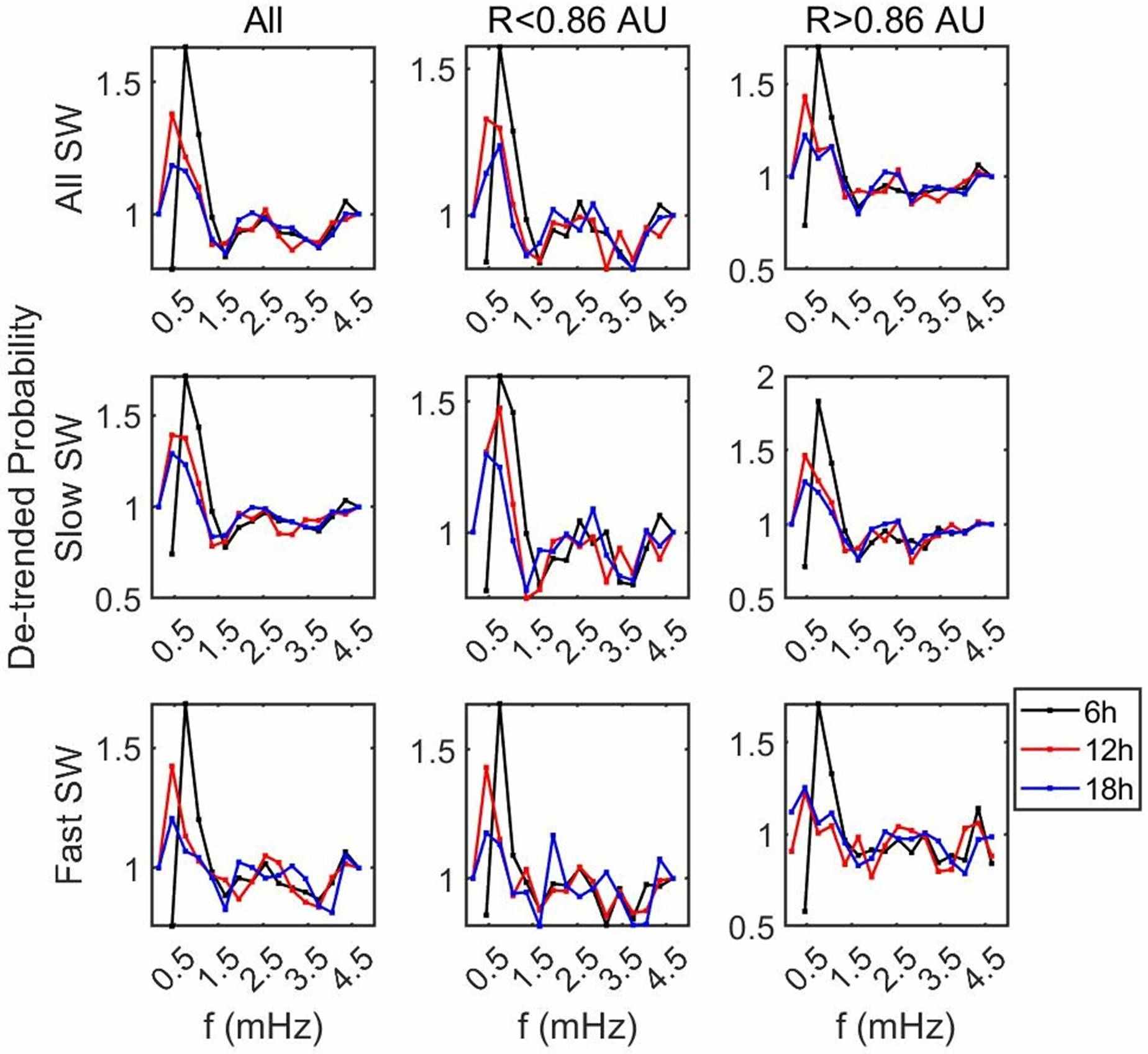}}     
\caption{De-trended probability of occurrence for PDS frequencies, where the distributions of the six-, twelve- and eighteen-hour interval are coloured with black, red and blue, respectively. From left to right: distributions for the entire sample, inner distances (R$<$0.86 AU) and near-L1 (R$>$0.86 AU), respectively. From top to bottom: distributions for the entire sample, slow (V$<$450 km/s) and fast (V$>$450 km/s) solar wind, respectively.}
\label{Fig2}
\end{figure}

The PDS frequency distributions are derived in the 0.2-5 mHz range binned into groups with $\Delta f$=0.3 mHz, which is the spectral window bandwidth for the MTM (see also Appendix A.1). As shown in figure \ref{FigA3} in the Appendix, PDS frequency distributions exhibit an increasing probability with increasing frequency, while for f$<$3 mHz, two local maxima appear at around 0.5 and 2.5 mHz, which is in agreement with previous studies \citep{Kepko2024,Katsavrias2025}. However, at f$>$3 mHz, the distributions exhibit an extremely abrupt increase. This feature is due to the fact that the Nyquist threshold favors the higher frequency PDS events. For reference, Figure \ref{FigA4} in the Appendix shows the frequency probability of occurrence without the use of the wavelet analysis. It can be seen that there is a similar increasing trend with increasing frequency, which after the application of the Nyquist filtering is magnified because many lower frequency PDS events do not pass the threshold. Another possible reason could be the presence of Alfv\'en waves or sudden density increases (e. g., shock, magnetic holes, etc.), which can introduce broad band artifacts in the spectrum, especially in the wavelet analysis \citep{Katsavrias2022}. 

In order to gain a clear view of the peaks in the frequency distribution, we perform a de-trending of the probability curves by dividing them with a moving average smoothing of 9 data points (half of the frequency bins). Figure \ref{Fig2} shows the de-trended PDS frequency distributions of the six-, twelve- and eighteen-hour interval analysis, in black, red and blue, respectively. As shown in the left panel of figure \ref{Fig2}, there are clear local maxima at $\approx$0.3-0.9 and $\approx$2.1-2.7 mHz. A secondary local peak at $\approx$4.3 mHz also appears (more prominent during the fast solar wind shown in the bottom left panel) but may be affected by the filtering, since it is very close to the cut-off frequency. Furthermore, the main peak at f$<$1 mHz shifts to lower frequencies for larger windows (0.6-0.9 and 0.3-0.6 mHz for the 6-hours and the 12- and 18-hours window, respectively). This is expected since larger windows can better resolve the very low frequencies. 

These maxima are in agreement with what has been found in 25 years data of ambient solar wind by \citet{Kepko2024}. The latter authors used Wind data and found broad occurrence enhancement in both the proton and n$_\alpha$/n$_p$ distributions with local maxima at $\approx$0.5, 1.5 and 2.7 mHz (see also Figure 9 in \citet{Kepko2024}). We note that the $\approx$1.5 mHz PDS are shown to be present during the descending phase and minimum of Solar cycles 23 and 24 (see also Figure 10 in \citet{Kepko2020}). However, the dataset used in this study spans only the rising phase and maximum of Solar cycle 25, therefore, our results are in agreement with the ones from Wind spacecraft. We finally note that the aforementioned results do not seem to differ with respect to the radial distance from the Sun.

\subsection{Distribution of PDS radial length scale ($L_R$)}

\begin{figure}
{\includegraphics[width=\hsize]{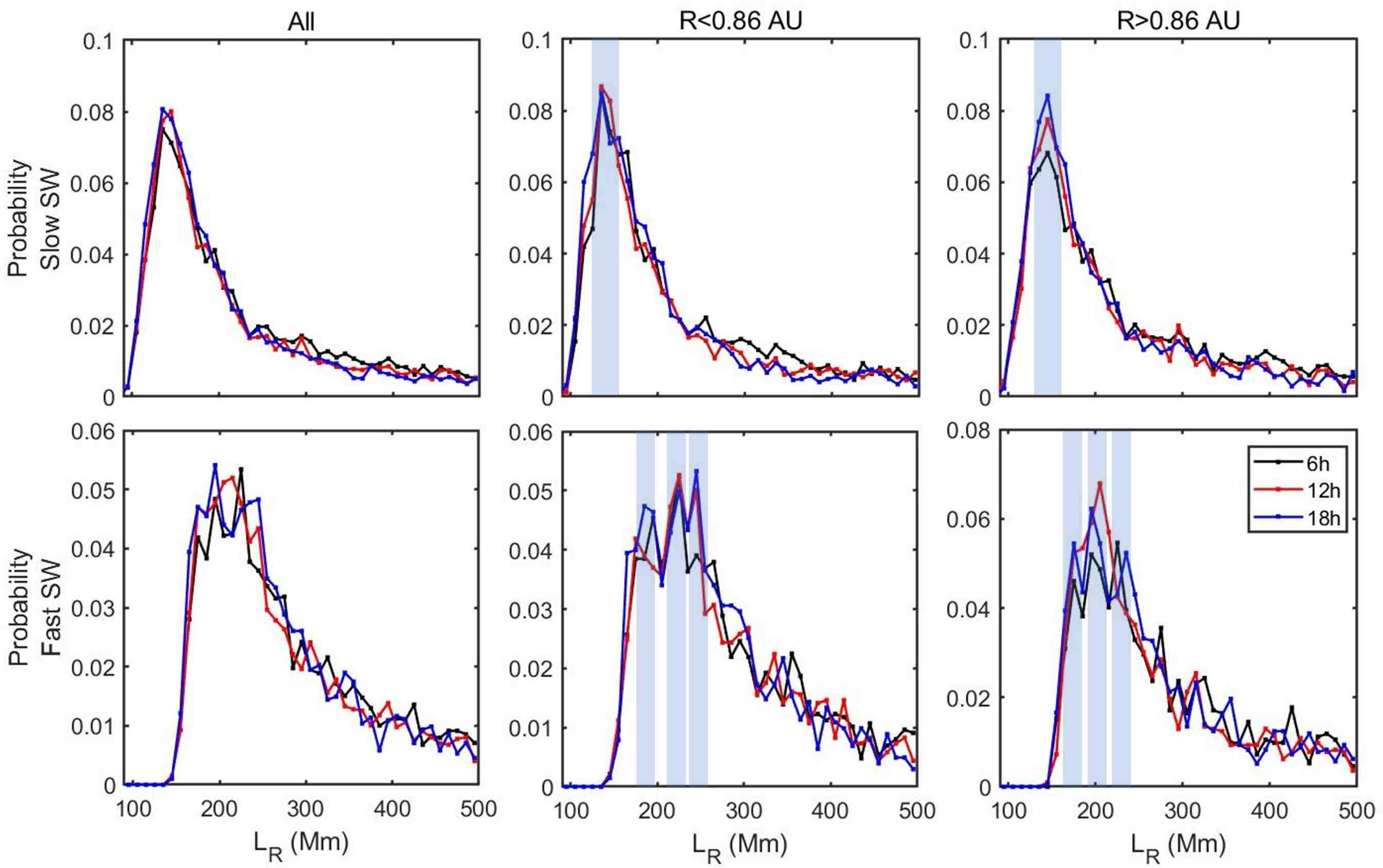}}     
\caption{Similar to figure \ref{Fig2} but for the PDS radial length scale (L$_R$). From top to bottom: distributions for the slow (V$<$450 km/s) and fast (V$>$450 km/s) solar wind, respectively. The light blue shaded rectangulars indicate the peaks of the distribution. }
\label{Fig3}
\end{figure}

Working in the frequency domain does not account for any compression or relaxation of the density structures that may have occurred in transit and may result in higher occurrence rate of periodicities in length scale \citep{Viall2008,Viall2009a}. To that end, we convert the frequencies to radial length scales using the mean solar wind radial velocity. In order to further account for the spacecraft velocity we calculate the radial length scale as: $ L_R = (V_R - V_{R,SC}) / f_{PDS} $, where $V_R$ and $V_{R,SC}$ are the mean solar wind and spacecraft radial velocity, in RTN coordinates, for each PDS event and $f_{PDS} $ the corresponding frequency. The correction for the spacecraft velocity is negligible near L1 but becomes important as SolO is near its perihelion. We note that we restrict the calculations to PDSs with f$<$3 mHz. The reason is that the abrupt increase in the occurrence of $>$3 mHz PDSs can artificially shift the true peaks in the radial length scale distribution to lower values. 

Figure \ref{Fig3} shows the $L_R$ distribution of the PDS events for the six-, twelve- and eighteen-hour interval analysis, in black, red and blue, respectively. During the slow solar wind segments, the scale distribution exhibits a peak at 130--140 Mm and 140--150 Mm, at inner distances (R$<$0.86 AU) and near L1, respectively. The distributions during the fast solar wind segments are more noisy. At inner distances (R$<$0.86 AU), there are local peaks at 180--190, 220--230 and 240--250 Mm, while the near-L1 distribution exhibits local peaks at 170--180, 190--200 and 220--230 Mm. These results indicate that, even though the frequency distribution is not affected with radial distance from the Sun, the PDS change size as they propagate.

\begin{figure}
{\includegraphics[width=\hsize]{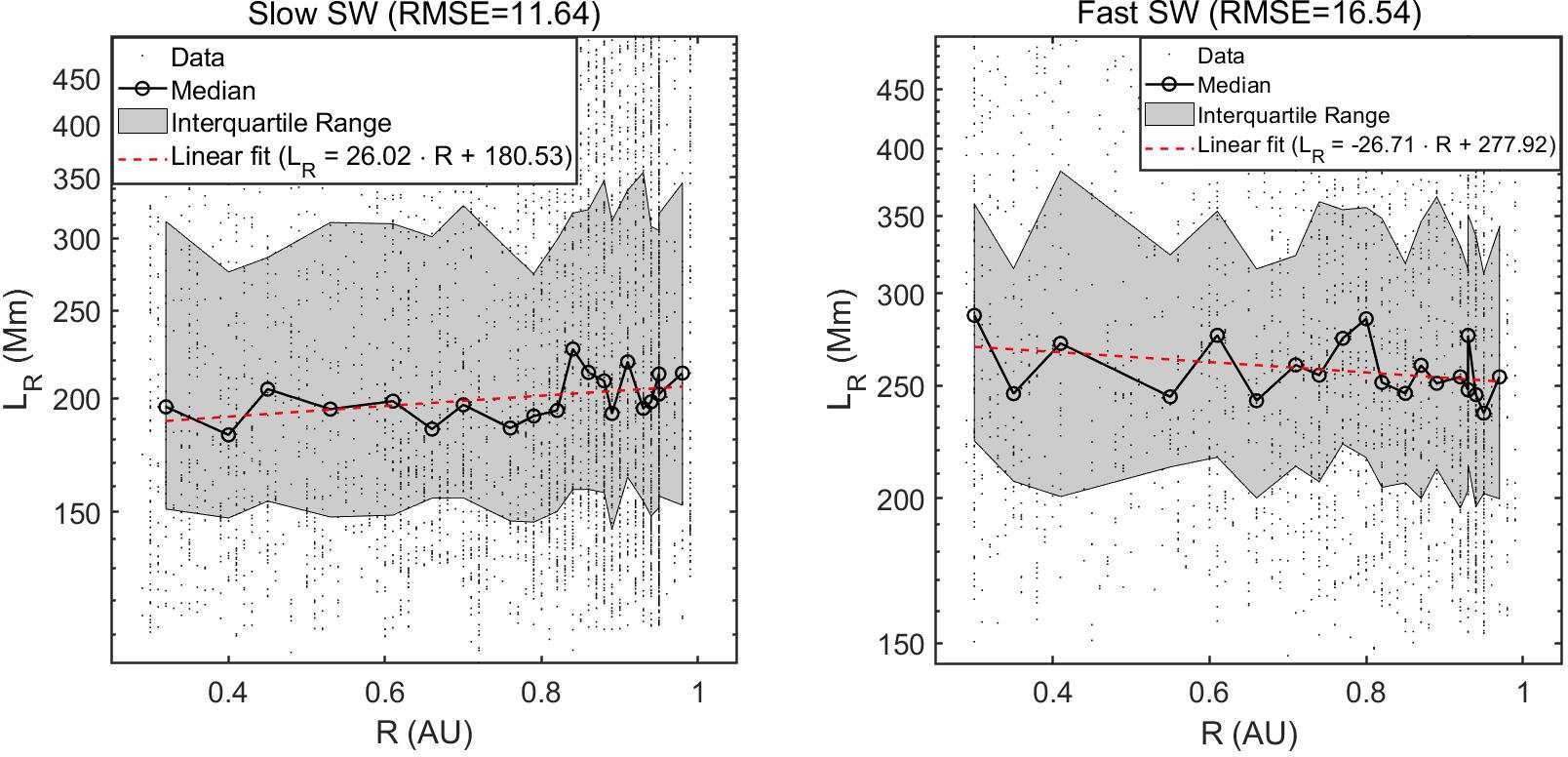}}     
\caption{Dependence of the PDS radial length scale during slow (left panel) and fast solar wind (right panel) on spacecraft radial distance for the 6-hour window. Grey dots correspond to the data and the solid black lines to the median of each R bin, respectively, while the grey shaded area is the inter-quartile range. The power law fit on the data is depicted with the solid red line, while the Root Mean Square Error (RMSE) is given on the top of each panel.}
\label{Fig4}
\end{figure}

To further test this scenario, we performed a binning of the PDSs $L_R$ with respect to the spacecraft radial distance (R). We note that the binning was performed in 20 bins that had variable lengths but contained the same amount of data points. Figure \ref{Fig4} shows the dependence of the median $L_R$ on the median R in each bin, along with the inter-quartile range for PDSs identified with the 6-hours window. As shown, there is an increase of $L_R$ with increasing distance from the Sun for the slow solar wind segments (left panel), while during fast solar wind segments, there is a decrease of $L_R$ with increasing distance from the Sun. The expansion and compression of the PDSs during slow and fast solar wind, respectively, is at a rate of $\approx$10\%. We note that the expansion and compression of the PDSs during slow and fast solar wind is similar if we consider different number of bins and/or the 12 and 18-hour windows, even though the corresponding absolute values of the slopes vary in the 15--30 range.

\section{Discussion and conclusions}
   
Using 3.5 years of Solar Orbiter measurements spanning the rising phase and maximum of Solar cycle 25 (2021--2024), We compiled and made publicly available an extensive list of PDS events in the 0.3 to 1 AU radial distances. The list includes 29891, 38478 and 37974 PDS events for the six-, twelve- and eighteen-hour interval, respectively, while the total duration of PDS events in our list is approximately 277, 414 and 443 days for the three window intervals. This means that the solar wind is comprised by PDSs at $\approx$22, 32 and 35\%, which is in agreement with previous statistical studies using Wind spacecraft \citep{Kepko2024}.  

Our results indicate that PDS expand as they advect with the ambient/slow solar wind. This is consistent with the scenario in which PDSs are formed at the Sun by processes involving magnetic reconnection and interchange reconnection in the solar corona \citep[e. g.,][]{Antiochos2011,Higginson2018,Reville2020}. On the other hand, PDSs detected in fast solar wind segments exhibit compression, which is indicative of their interaction with Stream Interaction Regions \citep{Kepko2019,Berriot2024,Katsavrias2025}. 

We note that this is the first study using such a large number of PDSs to statistically investigate their dependence on radial distance from the Sun. However, our conclusions are mostly qualitative and the exact rate of compression/expansion of the PDSs, as they advect with the solar wind, require a more detailed analysis. First, PDS frequency and radial length scale are dependent on the Solar cycle phase \citep{Kepko2020}, while our dataset covers the rising phase and maximum of Solar cycle 25, only. Second, and most important, we have no information about the proton and alpha particle density, since they were not publicly available at the time of this investigation. The contribution of protons is usually dominant in the total ion density, therefore we expect no significant changes in our list of events \citep{DiMatteo2024}. However, the $\alpha/p$ ratio is a crucial parameter to refine the list toward PDSs with a more likely Solar origin \citep{Kepko2016,Viall2021,DiMatteo2024}. In fact, coherent PDSs (shared periodic behaviour between proton density and $\alpha/p$ ratio) have been shown to exhibit very different properties from incoherent events \citep{Katsavrias2024}. These further considerations can better identify subgroups of PDSs and provide a more accurate determination of the rate of expansion/compression of such structures. Finally, Solar Orbiter spends much more time near-L1 rather than in smaller distances from the Sun (see also Figure \ref{FigA2} in the Appendix). Therefore, the combined use of Solar Orbiter and Parker Solar Probe (or similar missions) can provide a more robust picture of PDS events close to the Sun. This can be further investigated in the context of remote sensing observations. Particularly, the Polarimeter to Unify the Corona and Heliosphere (PUNCH) mission \citep{DeForest2025}, launched in March 2025 and specifically designed to image mesoscale density structures such as PDSs, can identify in situ observed structures as well as characterise the large scale context in which they occur.

\begin{acknowledgements}

This material is based upon work supported by the National Aeronautics and Space Administration through the completed Heliophysics Internal Scientist Funding Model Program. The authors thank the Solar Orbiter team for the use permission of the data and the ESA Solar Orbiter archive for making these data available (\url{https://soar.esac.esa.int/soar/}). CK received support from the ESA Archival Research Visitor Programme. S.D. was also supported by NASA Grant 80NSSC21K0459.

\end{acknowledgements}

\bibliographystyle{aa} 
 \bibliography{refs}

\begin{appendix} 

\section{Methods}

\subsection{Multitaper method} \label{MTMmethod}
We estimated the adaptive MTM power spectral density over running linearly detrended six-hour intervals using a time half-band width product $NW=2.5$ and number of tapers $K=4$ \citep{Slepian1978}. This choice of parameters resulted in a Rayleigh frequency of $f_{Ray}\approx$ 0.05 mHz and a spectral window bandwidth of $\approx2NWf_{Ray}\approx$ 0.3 mHz. The power spectral density was then normalized by the best background spectrum representation between those estimated via a maximum likelihood fitting method of a continuous background model over the logarithmically binned spectrum \citep[raw+BPL and bin+BPL; see details in][]{DiMatteo2021}. For the continuous background model, we use the pan-spectrum function \citep{Liu2020}:
\begin{equation}
    PSD(f)=cf^{-\beta} \left[1+ \left( \frac{f}{f_{b}}\right)\right]^{\beta-\gamma}
\end{equation}
which is a bending power law going from a slope of $\beta$ to $\gamma$ at a frequency break point, $f_{b}$. The background spectrum was estimated in the frequency range between $\approx$0.2 and $\approx$5 mHz, within the reliable frequency range as defined by \citet{DiMatteo2021}. 

\subsection{Continuous wavelet transform} \label{CWTmethod}
The analysis of a function in time, $F(t)$, into an orthonormal basis of wavelets is conceptually similar to the MTM. However, the latter is only localized in frequency, while the continuous wavelet transform (henceforward CWT), which is localized in frequency and time, allows for the local decomposition of nonstationary time series, providing a compact, two-dimensional representation. As most astrophysical time series are usually composed of sinusoidal-like oscillations, the most common mother wavelet used is the Morlet wavelet \citep{Morlet1982}, which consists of a complex plane wave modulated by a Gaussian. The equations are 
\begin{eqnarray}
\psi _o \left( n \right) = \frac{1}{\sqrt{4\pi} }\exp \left( i\omega _o n - \frac{n^2 }{2} \right) \\ 
 W_n^F \left( s \right) = \sqrt {\frac{{\delta t}}{s}}  \cdot \sum\limits_{n' = 0}^{N - 1} {F_{n'} } \psi _o^* \left( {\frac{{n' - n}}{s}\delta t} \right)
\label{morlet},
\end{eqnarray}
where $\omega_0$ is the dimensionless frequency, usually set to 6 to satisfy the admissibility condition, and $n$ is the dimensionless time \citep[see also][for further details]{Katsavrias2012,Katsavrias2021,Katsavrias2022}.

\section{Supplementary figures}
   
\begin{figure}
\includegraphics[width=\hsize]{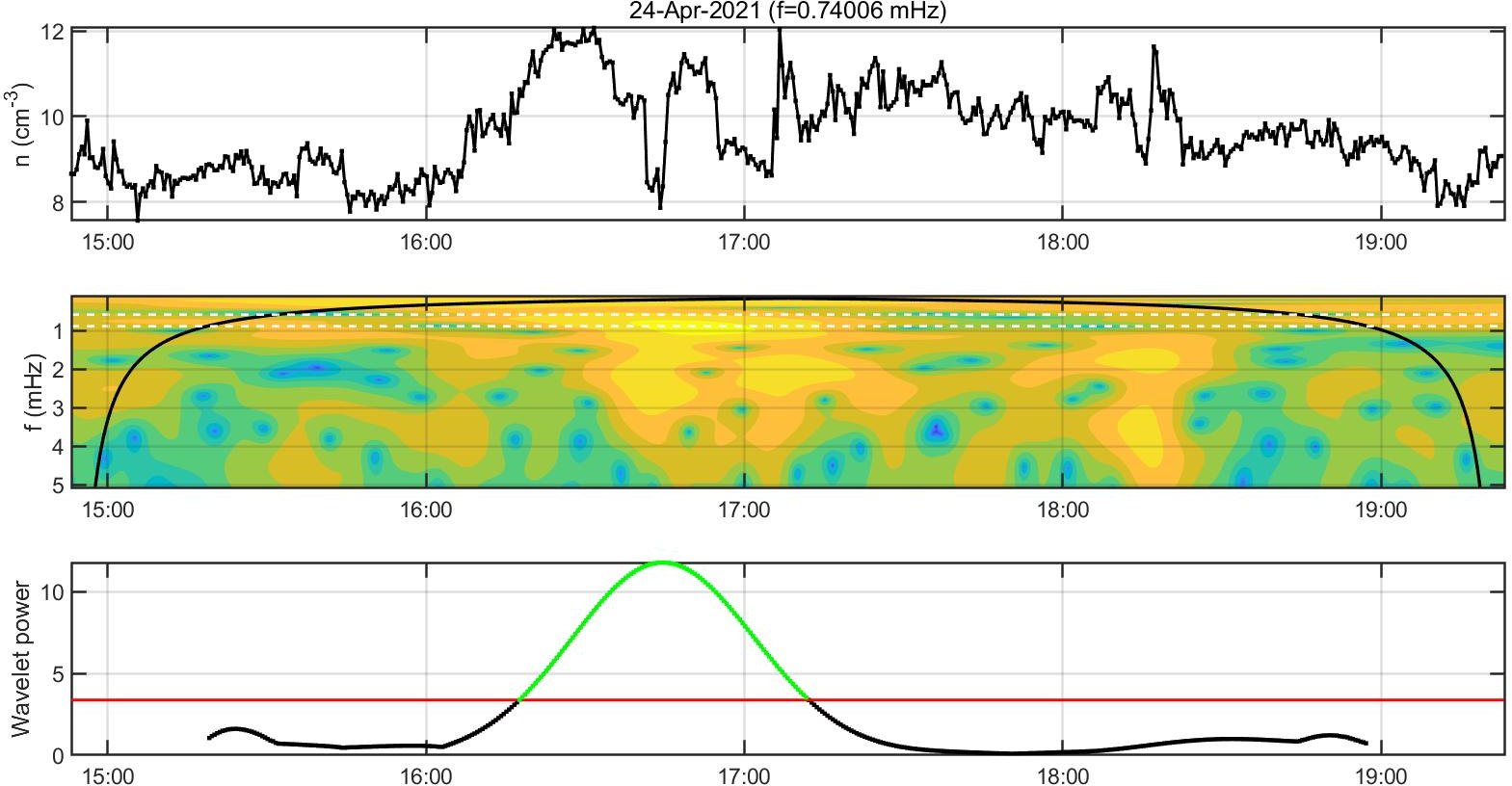}
\caption{Example of event detection using the continuous wavelet transform. Top panel: Time series of total ion density. Middle panel: Wavelet spectrum of the total ion density, where the horizontal white dashed lines correspond to the frequency range of interest (f$\pm0.15$ mHz) and the solid black line depict the cone of influence, where edge effects in the processing become important. Bottom panel: Average wavelet power in the f$\pm0.15$ mHz range. The horizontal red solid line corresponds to the 75th quantile of the power of the entire time period for the 6 hours window.}
\label{FigA1}
\end{figure}

\begin{figure}
\includegraphics[width=\hsize]{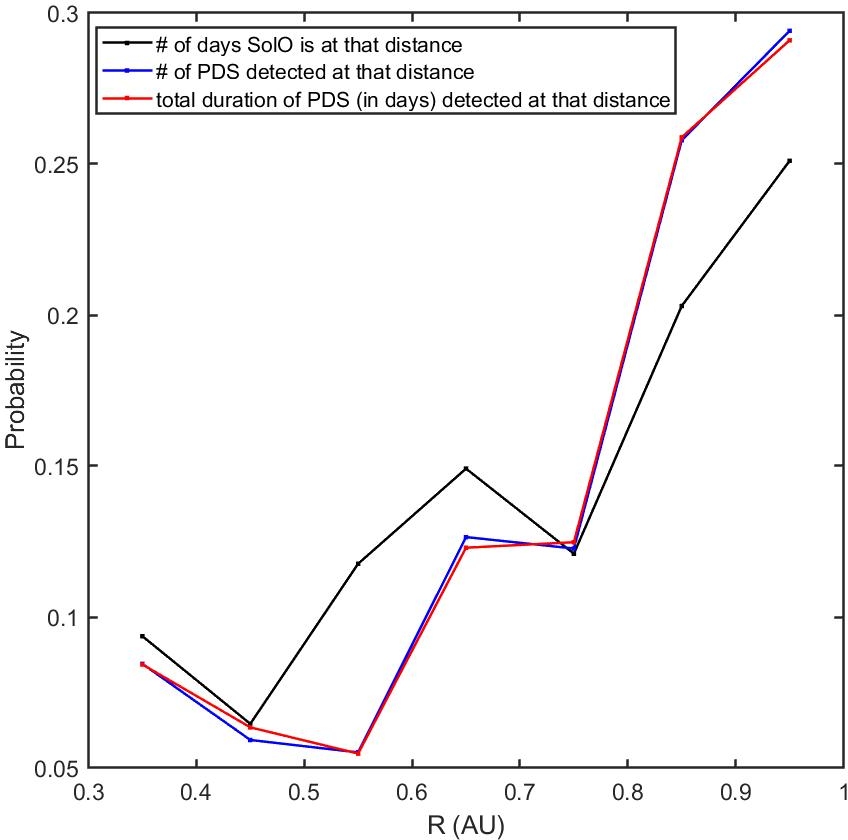}
\caption{Distributions of the number of days Solar Orbiter is at each R bin, the number of PDS detected in each R bin and total duration of the corresponding PDSs with black, blue and red lines, respectively.}
\label{FigA2}
\end{figure}

\begin{figure}
\includegraphics[width=\hsize]{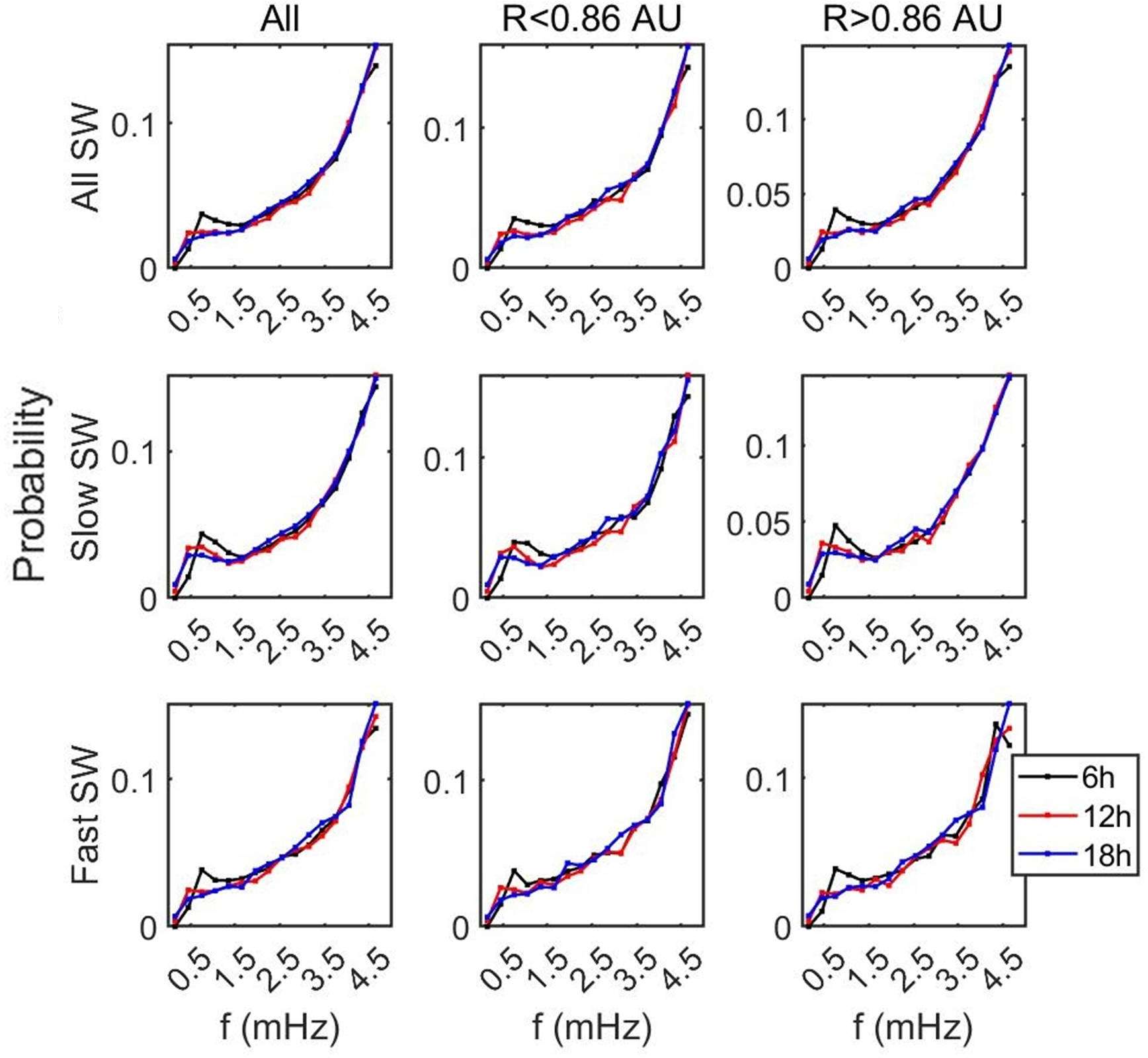}
\caption{Probability of occurrence for PDS frequencies, where the distributions of the six-, twelve- and eighteen-hour interval are coloured with black, blue and red lines, respectively. From left to right: distributions for the entire sample, inner distances (R$<$0.86 AU) and near-L1 (R$>$0.86 AU), respectively. From top to bottom: distributions for the entire sample, slow (V$<$450 km/s) and fast (V$>$450 km/s) solar wind, respectively. }
\label{FigA3}
\end{figure}

\begin{figure}
\includegraphics[width=\hsize]{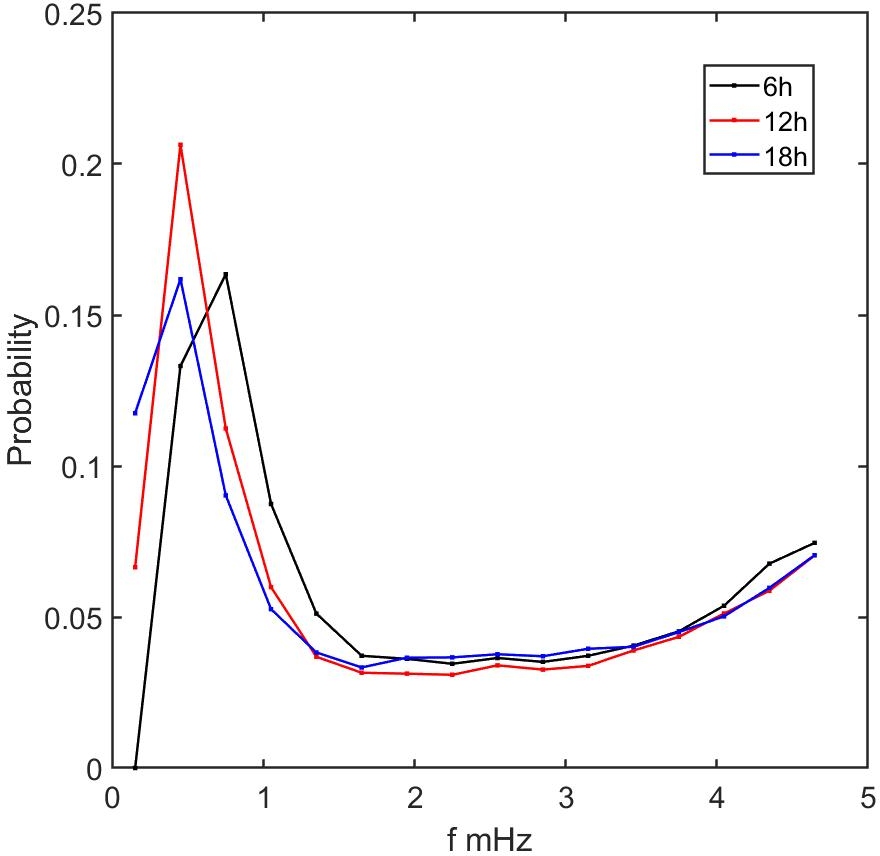}
\caption{Probability of occurrence for PDS frequencies detected with MTM method only, where the distributions of the six-, twelve- and eighteen-hour interval are coloured with black, blue and red lines, respectively.}
\label{FigA4}
\end{figure}

\end{appendix}

\end{document}